\documentclass[conference]{IEEEtran}
\makeatletter
\def\ps@headings{%
\def\@oddhead{\mbox{}\scriptsize\rightmark \hfil \thepage}%
\def\@evenhead{\scriptsize\thepage \hfil \leftmark\mbox{}}%
\def\@oddfoot{}%
\def\@evenfoot{}}
\makeatother
\usepackage{psfrag}

\usepackage{amssymb}
\usepackage[utf8]{inputenc}
\usepackage{graphicx}
\usepackage{float}
\usepackage{color, colortbl}
\usepackage{xcolor}
\usepackage{array}
\usepackage{multirow}
\usepackage{footnote}
\usepackage{cite}
\usepackage{threeparttable}

\makesavenoteenv{tabular}

\begin{document}
%opening
 \title{Fairness in Collision-Free WLANs}

%A more simple output, useful when involving people from different affiliations
  \author{
      \IEEEauthorblockN{Luis Sanabria-Russo\IEEEauthorrefmark{0}, Jaume Barcelo\IEEEauthorrefmark{0}, Boris Bellalta\IEEEauthorrefmark{0}}\\
      \IEEEauthorblockA{\IEEEauthorrefmark{0}Universitat Pompeu Fabra, Barcelona, Spain
      \\\{luis.sanabria, jaume.barcelo, boris.bellalta\}@upf.edu}
  }

%This is the style of three columns, as indicated in IEEEtran
% \author{\IEEEauthorblockN{Luis Sanabria-Russo}
%  \IEEEauthorblockA{Department of Information\\
%  and Communications Technologies\\
%  Universitat Pompeu Fabra\\
%  Barcelona, Spain\\
%  Email: luis.sanabria@upf.edu}
%  \and
%  \IEEEauthorblockN{Jaume Barcelo}
%  \IEEEauthorblockA{Department of Information\\
%  and Communications Technologies\\
%  Universitat Pompeu Fabra\\
%  Barcelona, Spain\\
%  Email: cristina.cano@upf.edu}
%  \and
%  \IEEEauthorblockN{Boris Bellalta}
%  \IEEEauthorblockA{Department of Information\\
%  and Communications Technologies\\
%  Universitat Pompeu Fabra\\
%  Barcelona, Spain\\
%  Email: boris.bellalta@upf.edu}}

\maketitle

\begin{abstract}

\boldmath CSMA/ECA is a contention protocol that makes it possible to construct a collision-free schedule by using a deterministic backoff after successful transmissions. In this paper, we further enhance the CSMA/ECA protocol with two properties that allows to fairly accommodate a large number of contenders in a collision-free schedule. The first property, called \emph{hysteresis}, instructs the contenders not to reset their contention window after successful transmissions. Thanks to hysteresis, the protocol sustains a high throughput regardless of the number of contenders. The second property, called \emph{fair-share}, preserves fairness when different nodes use different contention windows. We present simulations results that evidence how these properties account for performance gains that go even further beyond CSMA/CA.

\end{abstract}


\section{Introduction} \label{introduction}
  IEEE $802.11$ networks use a shared medium to establish communication among nodes. Carrier Sense Multiple Access with Collision Avoidance (CSMA/CA) is the protocol in charge of coordinating access to the wireless medium in order to avoid simultaneous transmissions by different nodes. If two or more nodes attempt transmission at the same time, a \emph{collision} occurs and the resulting transmission is discarded by the receivers.

Carrier Sense Multiple Access with Enhanced Collision Avoidance (CSMA/ECA)~\cite{CSMA_ECA} was introduced as an enhancement to CSMA/CA. It is capable of achieving a collision-free state by making very simple changes on the way CSMA/CA behaves: choosing a deterministic backoff after successful transmissions. CSMA/ECA preserves backward compatibility with CSMA/CA (details in~\cite{CSMA_ECA}~and~\cite{HE}), which is paramount for the coexistence and progressive adoption of the protocol. 

The performance evaluation for CSMA/ECA has been presented in~\cite{E2CA_performance}. Nevertheless, to the best of our knowledge this is the first work that introduces further enhancements to the protocol, making it possible to allocate a larger number of contenders and achieve greater throughput than CSMA/CA while providing throughput fairness to all users. This is the first step towards the construction of a totally distributed MAC protocol with better performance than the current standard as a consequence of its collision-free operation.

\section{Background}

Time in WLANs is slotted, and each slot can be classified as empty, successful or collision (accounting for no transmission, successful transmission or collision, respectively).

In CSMA/CA, each contender attempting to transmit a packet chooses a backoff counter $B\in[0,CW(k)-1]$ randomly, where $k\in[0,\ldots,m]$ is the \emph{backoff stage} and $CW(k)=2^{k}CW_{\min}$ is the contention window, with $CW_{\min}$ its minimum value. Each passing empty slot decrements $B$ by one; when the backoff counter reaches zero, the contender will attempt transmission. The success of the transmission attempt is only confirmed by the reception of an acknowledgement (ACK) frame from the receiver, otherwise a collision is assumed. If that is the case, each contender involved in the collision doubles its contention window by incrementing its backoff stage and the packet is retransmitted. If the transmission is successful, the sender resets its contention window to the minimum value ($CW(0)=CW_{\min}$).

CSMA/ECA achieves less collisions and outperforms CSMA/CA in most typical scenarios (see~\cite{HE} and \cite{E2CA_performance}). The only difference with CSMA/CA is that a deterministic \mbox{backoff}~\mbox{$B_{\mbox{\scriptsize{d}}}=CW_{\min}/2$} is chosen after each successful transmission. This choice makes it possible for CSMA/ECA to fairly coexist with CSMA/CA~\cite{CSMA_ECA}. Furthermore, the maximum number of contenders that can be accommodated in a collision-free fashion in CSMA/ECA is equal to the deterministic backoff used after successful transmissions $B_{\mbox{\scriptsize{d}}}$.
%From here on $B_{\mbox{\scriptsize{d}}} = N_{\max}^{\mbox{\scriptsize{cf}}}$, where~$N_{\max}^{\mbox{\scriptsize{cf}}}$ is defined in~(\ref{eq:capacity}) as the \emph{collision-free constraint} and represents the maximum number of nodes participating in the contend for transmission able to achieve a collision-free state.

%\begin{equation} \label{eq:capacity}	
%	N_{\max}^{\mbox{\scriptsize{cf}}} = \lceil{CW_{\min}/2}\rceil
%\end{equation}

In a scenario where the number of contenders, $N$, is not larger than the deterministic backoff $B_{\mbox{\scriptsize{d}}}$, eventually all contenders will be able to pick different transmission slots, therefore achieving a collision-free state.

\begin{figure}[htbp]
  \centering
  \includegraphics[width=0.95\linewidth]{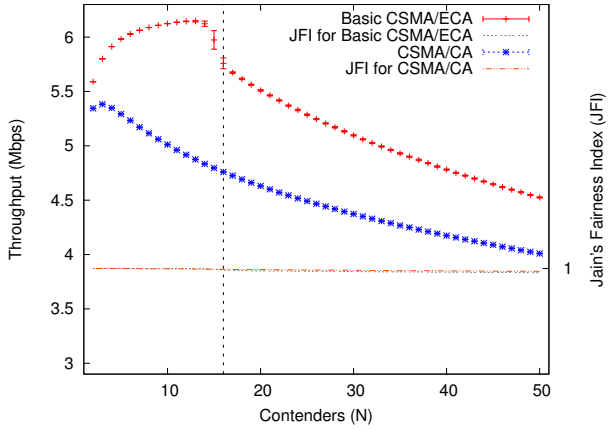}
  \caption{The throughput is CSMA/ECA decreases when the number of contenderes~$N$~exceeds~$B_{\mbox{\scriptsize{d}}}$, which is the maximum number of contenders that can be allocated in a collision-free fashion.
  \label{fig:throughput}}
\end{figure}

When the system is overcrowded, $N>B_{\mbox{\scriptsize{d}}}$, CSMA/ECA suffers a decrease in throughput due to the fact that it is impossible to reach a collision-free operation. This effect can be seen in Figure~\ref{fig:throughput}, where $CW_{\min}=32$ and $B_{\mbox{\scriptsize{d}}}=16$.

%as seen in Figure~\ref{fig:throughput}. This effect is caused by the fact that it is impossible to reach collision-free operation.

%This effect is caused by collisions originated by $N-B_{\mbox{\scriptsize{d}}}$ contenders forced to generate a random backoff counter and attempting transmission on slots previously picked by $N_{\max}^{\mbox{\scriptsize{cf}}}$ nodes using a deterministic backoff.

%Furthermore, as $N-N_{\max}^{\mbox{\scriptsize{cf}}}$ nodes are unable to successfully transmit, collisions force the nodes that chose a deterministic backoff, to switch to a random one. 

The outcome is a mixed system composed of contenders using either deterministic or random backoff counters. Note that the throughput in CSMA/ECA is greater than CSMA/CA's for any number of contenders (Figure~\ref{fig:throughput}).

%The throughput degradation depicted in Figure~\ref{fig:throughput} when $N>16$, is a consequence of the great number of collisions resulting from this behavior. 

\section{A descentralized and fair CSMA/ECA} \label{csmae2ca}
  %Nodes will double $CW(k)$ after collisions and reset it ($CW(k)=CW_{min}$) upon each transmission success, augmenting the collision probability. 
Because CSMA/ECA is totally distributed, the number of nodes ($N$) is unknown to all contenders. In the following we introduce a mechanism able to reach collision-free operation without knowledge of $N$, even for a large number of contenders.
%However the backoff stage of each station provides a hint of the level of contention.

%Therefore, $CW(k)$ is used to relate collisions to the number of users in the system.

%$N$ is unknown to all contenders, so $CW(k)$ is the only indicator for each node of the current state of the system.

To make it possible to achieve a collision-free state when the system is overcrowded, we instruct nodes not to reset $CW(k)$ after successful transmissions, and pick a deterministic backoff $B_{\mbox{\scriptsize{d}}}=CW(k)/2$. This is called \emph{hysteresis} from here on. 

%and modify the collision-free constraint to~$N_{\max}^{\mbox{\scriptsize{cf}}}=CW(k)/2$

%Hysteresis forces nodes to \emph{stick} to the value of the current backoff stage, $k$; resulting in a deterministic backoff greater than the minimum contention window.

%This measure leads to a collision-free state while $N\leq B_{\mbox{\scriptsize{d}}}$.

\begin{figure}[htbp]
  \centering
  \includegraphics[width=0.95\linewidth]{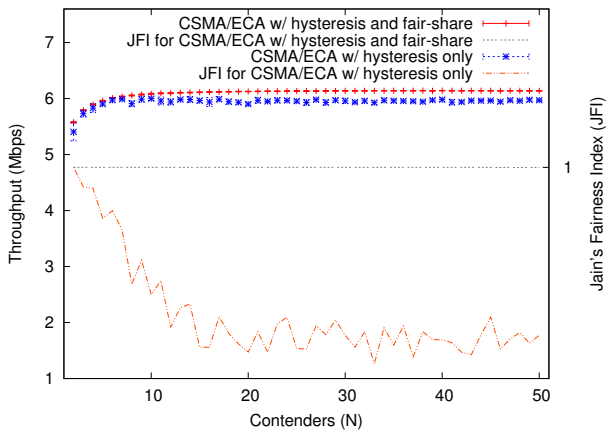}
  \caption{Throughput and Jain's Fairness Index when implementing hysteresis and fair-share in CSMA/ECA
  \label{fig:fairShare}}
\end{figure}

Hysteresis produces deterministic backoffs that are larger than $CW_{\min}/2$, thus making it possible to allocate more contenders in a collision-free fashion. Contenders may be in different deterministic backoff stages, which provokes some nodes to access the channel more often than others. This fairness issue, that can be observed in Figure~\ref{fig:fairShare}, is averted with \emph{fair-share}. The concept of fair-share, was first introduced by Fang et al. in~\cite{L_MAC2}.
%Having a greater collision-free constraint means that more nodes are able to achieve a collision-free state. 

Fair-share consist in allowing each contender to send $2^{k}$ packets at every transmission, making sure that contenders with longer backoff are compensated proportionally.

Figure~\ref{fig:fairShare}, depicts how CSMA/ECA with hysteresis and fair-share achieves greater throughput than CSMA/ECA with hysteresis only, maintaining a collision-free state while being fair (Jain's Fairness Index~\cite{JFI}~(JFI) equal to $1$), for any number of contenders.

This work evaluates the performance of CSMA/ECA when implementing the concept in a customized C++ simulator.

\section{Evaluation}
%Other performance evaluations, like a semi-analytical framework modelling the enhanced collision avoidance mechanism and comparing it with other access schemes (like Basic Access and RTS/CTS), are provided in~\cite{E2CA_performance}. Nevertheless, to the best of our knowledge this is the first evaluation of resilience and fair-share in CSMA/ECA.

Implementation is performed on a customized version of the COST~\cite{COST}~simulator. The system was set to be under saturation (nodes always have packets to transmit) during a period of ten seconds at a maximum throughput of $11$Mbps. The number of contenders ranges from $2$ to $50$ and a hundred simulations are performed for each number of contenders. Further MAC-related parameters as well as the code for the whole CSMA/ECA implementation can be found in~\cite{sim:parameters}.

Figure~\ref{fig:throughput} and Figure~\ref{fig:fairShare} are results derived from the evaluation platform with $95\%$ confidence intervals. Note that the confidence intervals are so small that can hardly be appreciated in the figure.

\section{Future directions} \label{future}
  %CSMA/ECA is the basic idea, but there are many issues to investigate in the future. One of them, is to make it adaptive to a variable number of nodes without fairness between nodes. Open topics include how to provide traffic differentiation (i.e. Quality of Service) in top of CSMA/ECA, or how to adapt novel features such as Multi-packet Transmission / Reception and channel bonding on top of it. 

%As it is shown in~\cite{E2CA_performance}, the combination of CSMA/ECA with the Auto Rate Fallback mechanism to select the transmission rates provides a huge gain in the network performance as, once collisions are removed, they do not interfere with the ARF operation. If similar gains can be obtained when combining it with the previously mentioned mechanisms are still open challenges.

To produce a throughout analysis of CSMA/ECA, more evaluations need to be carried out under non-saturated conditions. Further enhancements include the reset of the backoff stage when the transmission queue is empty and to determine what is its impact on the overall performance of the protocol.

%Further enhancements to CSMA/ECA include the reset of the backoff stage when the transmission queue is empty as well as performance tests in non-saturated conditions. This introduces new challenges related to convergence time and delay which need to be leveraged.

Also, future development will be focused on implementing CSMA/ECA in cheap commodity hardware~\cite{WMP}. Doing so will open the door for evaluation under more realistic scenarios as well as provide insight on different communication aspects, for example those regarding channel errors, delay, synchronization, coexistence with other access protocols and real network traffic. 

%Nevertheless, this is not a trivial task given that in requires flexible network interface hardware which is not commonly provided by manufacturers. 

\section*{Acknowledgments}
The authors would like to thank Azadeh Faridi for her insightful comments and contributions.
  
\bibliographystyle{Classes/IEEEtran}
\bibliography{IEEEabrv,ref}
  
\end{document}